\documentstyle[preprint, epsfig, aps]{revtex}
\tightenlines

\begin{document}
\draft

\title{ Physical Limits on the Notion of Very Low Temperatures }

\author{Juhao Wu and A. Widom\\
Department of Physics, Northeastern University, Boston,
Massachusetts 02115}

\maketitle
\begin{abstract}

Standard statistical thermodynamic views of temperature fluctuations 
predict a magnitude $(\sqrt{<(\Delta T)^2>}/T)\approx\sqrt{(k_B/C)}$ for 
a system with heat capacity $C$. The extent to which low temperatures 
can be well defined is discussed for those systems which obey the 
thermodynamic third law in the form  $\lim_{(T\to 0)}C=0$. Physical limits 
on the notion of very low temperatures are exhibited for simple systems. 
Application of these concepts to bound Bose condensed systems are explored,
and the notion of bound Boson superfluidity is discussed in terms of the 
thermodynamic moment of inertia.

\end{abstract}
\medskip

\section{Introduction} 
A problem of considerable importance in low temperature physics 
concerns physical limitations on how small a temperature can be well 
defined in the laboratory[1]. In what follows, we shall consider 
temperature fluctuations which define a system temperature ``uncertainty'' 
$$
\delta T=\sqrt{<(T_s-<T_s>)>^2}\ , \eqno(1.1)
$$
whenever a finite system at temperature $T_s$ is in thermal 
contact with a large reservoir at bath temperature $T$. 
Only on the thermodynamic average do we expect 
the system temperature to be equal to the bath 
temperature; i.e.  $<T_s>\approx T$. The fluctuation from this 
average result (quoted in the better textbooks discussing statistical 
thermodynamics[2]), has the magnitude  
$$
\Big({\delta T\over T}\Big)\approx \sqrt{k_B\over C}, \eqno(1.2)
$$
where $C$ is the heat capacity of the finite system, and $k_B$ is 
Boltzman's constant. Since $C$ is an extensive thermodynamic quantity, 
one expects the usual small fluctuation in temperature 
$(\delta T/T) \propto (1/\sqrt{N})$ in the thermodynamic limit 
$N\to \infty $, where $N$ is the number of microscopic particles. 
However, in low temperature physics (for systems with finite values for 
$N$), temperature fluctuations are by no means required to be negligible.

For those finite sized systems which obey the thermodynamic third law  
$$
\lim_{T\to 0}C=0, \eqno(1.3)
$$ 
one finds from Eqs.(1.2) and (1.3) that 
$$
\Big({\delta T\over T}\Big)\to \infty \ \ 
{\rm as}\ \ T\to 0 \ \ {\rm with}\ \  N<\infty . \eqno(1.4)
$$
Eq.(1.4) sets the limits on what can be regarded as the ultimate 
lowest temperatures for finite thermodynamic systems; 
i.e. the temperature must at least obey $\delta T<<T$.

In Sec.II, the theoretical foundations for Eq.(1.4) will be discussed.  
In brief, the micro-canonical entropy of a thermodynamic system is given 
by 
$$
S(E)=k_B ln\Gamma (E), \eqno(1.5)
$$
where $\Gamma (E)$ is the number of system quantum states with energy 
$E$. The micro-canonical entropy defines the {\it system temperature} 
$T_s$ via 
$$
\Big({1\over T_s}\Big)=\Big({dS\over dE}\Big). \eqno(1.6)
$$
The {\it thermal bath temperature} $T$, which is not quite the system 
temperature $T_s$, enters into the canonical free energy 
$$
F(T)=-k_BT\ lnZ(T), \eqno(1.7)
$$
where the partition function is defined as 
$$
Z(T)=Tr\Big(e^{-H/k_BT}\Big)
=\sum_E \Gamma (E)e^{-E/k_B T}. \eqno(1.8)
$$

The probability distribution for the energy of the system, when in 
contact with a thermal bath at temperature $T$, is given by 
$$
P(E;T)=\Big({\Gamma (E)\over Z(T)}\Big)exp\Big({-E\over k_BT}\Big)
=exp\Big({F(T)-E+TS(E)\over k_BT}\Big), \eqno(1.9)
$$
as dictated by Gibbs. Thus, the temperature $T$ (of the thermal bath) 
does not fluctuate while system energy $E$ does fluctuate 
according to the probability rule of Eq.(1.9). On the other hand, 
the system temperature $T_s(E)$ in Eq.(1.6) depends on the system energy 
and thereby fluctuates since $E$ fluctuates. It is only for the energy 
$E^*$ of maximum probability $Max_{E} P(E;T)=P_{max}=P(E^*;T)$ 
that the system temperature is equal to the bath temperature 
$T_s(E^*)=T$. If the energy probability distribution is spread 
out at low thermal bath temperatures, then temperature fluctuations 
are well defined in the canonical ensemble of Gibbs.

In Sec.III, temperature fluctuations are illustrated using the example of 
black body radiation in a cavity of volume $V$. For this case, it 
turns out that the thermal wave length $\Lambda_T$ of the radiation 
in the cavity,  
$$
\Lambda_T=\Big({\hbar c\over k_BT}\Big), \eqno(1.10)
$$
must be small on the scale of the cavity length $V^{1/3}$; i.e. 
$\Lambda_T<<V^{1/3}$. For example, in a cavity of volume $V\sim 1\ cm^3$, 
the lowest temperature for the radiation within the cavity is of order 
$T_{min}\sim 1\ ^oK$. It is of course possible to cool the conducting metal 
walls of a cavity with a length scale $\sim 1\ cm$ to well below 
$1\ ^oK$. However, this by no means implies that the radiation within 
the cavity can have a temperature well below $1\ ^oK$. The point is that at 
temperatures $T_s<1\ ^oK$, there are perhaps only a few photons (in total) in 
the cavity. The total number of photons are far too few for the cavity 
radiation system temperature to be well defined. 

In Sec.IV, a confined system of ${\cal N}$ atoms 
obeying ideal gas Bose statistics is discussed. Such systems 
can be Bose condensed, and are presently (perhaps) 
the lowest temperature systems available in laboratories. In the 
quasi-classical approximation, the free energy is computed in Sec.V. 
Questions concerning bounds on ultra-low temperatures are explored. 
Whether or not such Bose condensed atoms can exhibit superfluid 
behavior is discussed in Sec.VI. The superfluid and normal fluid 
contributions to the moment of inertia are computed. 
In the concluding Sec.VII, another simple system with fluctuation limits 
on ultra-low temperatures will be briefly discussed. 

\section{Theoretical Foundations}
Let $\phi (E)$ denote some physical quantity which depends on the 
energy $E$ of a physical system. If the system is in contact with a 
thermal bath at temperature $T$, then the thermodynamic average may be 
calculated from 
$$
<\phi >=\sum_E P(E;T)\phi (E), \eqno(2.1) 
$$ 
where the probability $P(E;T)$ has been defined in Eq.(1.9). Using the 
``summation by parts''[3,4] formula 
$$
\sum_E {\partial \over \partial E}\Big(P(E;T)\phi (E)\Big)=0, \eqno(2.2)
$$
i.e. with a strongly peaked $P(E;T)$ 
$$
-\sum_E P(E;T)\Big({d\phi (E)\over dE}\Big)=
\sum_E \phi (E)\Big({\partial P(E;T)\over \partial E}\Big), \eqno(2.3)
$$
one finds 
$$
-k_B\Big<{d\phi (E)\over dE}\Big>=
k_B\sum_E \phi (E)\Big({\partial P(E;T)\over \partial E}\Big). \eqno(2.4)
$$
Employing Eqs.(1.6) and (1.9), 
$$
k_B\Big({\partial P(E;T)\over \partial E}\Big)=
\Big({1\over T_s(E)}-{1\over T}\Big)P(E;T). \eqno(2.5)
$$
Eqs.(2.4) and (2.5) imply the central result of this section 
$$
-k_B\Big<{d\phi \over dE}\Big>=
\Big<\Big({1\over T_s}-{1\over T}\Big)\phi \Big>. \eqno(2.6)
$$

If we choose $\phi (E)=1$, then Eq.(2.6) reads 
$$
\Big<{1\over T_s}\Big>={1\over T}\ ; \eqno(2.7)
$$
i.e. on average, the reciprocal of the system temperature is equal to 
the reciprocal of the thermal bath temperature. Thus, with fluctuations 
from the mean 
$$
\Delta \phi =\phi -<\phi >, \eqno(2.8)
$$
$$
\Delta \Big({1\over T}\Big)={1\over T_s}-\Big<{1\over T_s}\Big>
={1\over T_s}-{1\over T}, \eqno(2.9)
$$
Eq.(2.6) reads 
$$
-k_B\Big<{d\phi \over dE}\Big>=
\Big<\Delta\Big({1\over T}\Big)\Delta \phi \Big>. 
\eqno(2.10)
$$
If we choose in Eq.(2.10) the function $\phi $ to be  
$$
\phi =\Big({1\over T_s}\Big)\ \ {\rm and}\ \ 
-\Big({d\phi \over dE}\Big)={1\over T_s^2}\Big({dT_s\over dE }\Big)
=\Big({1\over T_s^2 C}\Big) \eqno(2.11) 
$$
(where $C=(dE/dT_s)$ is the system heat capacity), then 
$$
\Big<\Delta \Big({1\over T}\Big)^2\Big>
=\Big<{1\over T_s^2}\Big({k_B\over  C}\Big)\Big>. \eqno(2.12)
$$
The standard Eqs.(1.1) and (1.2) follow from the more precise 
Eqs.(2.7) and (2.12) in the limit of small temperature fluctuations; 
i.e. 
$$
<(\Delta T)^2>\approx \Big({k_B T^2\over C}\Big)\ \ \ 
{\rm if}\ \ \ \delta T=\sqrt{<(\Delta T)^2>}<<T. \eqno(2.13)
$$

The condition $\delta T<<T$ is required in order that the canonical 
thermal bath temperature be equivalent to the micro-canonical system 
temperature. If the micro-canonical and canonical temperatures are not 
equivalent, then the statistical thermodynamic definition of temperature 
would no longer be unambiguous. This raises fundamental questions as to the 
physical meaning of temperature. The view of this work is that in an ultra-low 
temperature limit, whereby $\delta T\sim T$ for sufficiently small $T$, the 
whole notion of system temperature is undefined, although the notion of a 
thermal bath temperature retains validity.

\section{Black Body Radiation Example}
The heat capacity of black body radiation in a cavity of volume 
$V$ with the walls of the cavity at temperature T is given by[5] 
$$
C({\rm Black\ Body})=k_B\Big({4\pi^2 \over 15}\Big)
\Big({V \over \Lambda_T^3}\Big), \eqno(3.1)
$$ 
where $\Lambda_T$ is given by Eq.(1.10). From Eqs.(1.2) and (3.1) 
it follows that the radiation temperature of a black body cavity of 
volume $V$ is 
$$
\Big({\delta T({\rm Black\ Body})\over T}\Big)\approx 
\sqrt{\Big({15 \over 4\pi^2 }\Big)\Big({\Lambda_T^3\over V}\Big)}
\approx 0.6\sqrt{\Big({\Lambda_T^3\over V}\Big)}. \eqno(3.2)
$$
In order to achieve a well defined radiation temperature inside the 
cavity, $\delta T({\rm Black\ Body})$ must be small on the scale of 
$T$ or equivalently $\Lambda_T<<V^{1/3}$. As stated in Sec.1, 
this implies a minimum temperature of $T_{min}\sim 1\ ^oK$ for a cavity of 
$V\sim 1\ cm^3$.

\section{Confined Ideal Bose Gas}

The grand canonical free energy of an ideal Bose gas is determined 
by the trace[6] 
$$
\Xi(T,\mu )=k_BT\ tr\ ln\Big(1-e^{(\mu -h)/k_BT}\Big), \eqno(4.1)
$$
where $h$ is the one Boson Hamiltonian and 
$$
d\Xi=-{\cal S}dT-{\cal N}d\mu  \eqno(4.2)
$$
determines the number of Bosons ${\cal N}$. If the one Boson partition 
function is defined as 
$$
q(T)=tr\Big(e^{-h/k_BT}\Big), \eqno(4.3)
$$
then the free energy obeys 
$$
\Xi(T,\mu )=-k_BT\sum_{n=1}^\infty \Big({1\over n}\Big)
q\Big({T\over n}\Big)e^{n\mu / k_BT}. \eqno(4.4)
$$
The mean number of Bosons is 
$$
{\cal N}(T,\mu )=\sum_{n=1}^\infty q\Big({T\over n}\Big)e^{n\mu / k_BT}, 
\eqno(4.5)
$$
and the statistical entropy is given by
$$
{\cal S}(T,\mu )=
-\Big({\Xi(T,\mu )+\mu {\cal N}(T,\mu )\over T}\Big)+
k_BT\sum_{n=1}^\infty  \Big({1\over n^2}\Big)
q^\prime \Big({T\over n}\Big)e^{n\mu / k_BT},  \eqno(4.6)
$$
where $q^\prime (T)=\{dq(T)/dT\}$.

Of considerable theoretical[7,8] experimental[9,10,11] interest is the 
bound Boson in an anisotropic oscillator potential,  
$$
h=-\Big({\hbar^2 \over 2M}\Big)\nabla^2+{\Big({1\over 2}\Big)}M
{\bf r\cdot }{\hat{\omega }^2}{\bf \cdot r}-
\Big({\hbar \over 2}\Big)tr(\hat{\omega }) 
\eqno(4.7)
$$
where 
$$
\hat{\omega }^2=
\pmatrix{ \omega_1^2 & 0 & 0 \cr
0 & \omega_2^2 & 0 \cr
0 & 0 & \omega_3^2 }. \eqno(4.8)
$$
Eqs.(4.3) and (4.7) imply 
$$
q(T)=\prod_{j=1}^3\Big({1\over 1-e^{-\hbar \omega_j/k_BT}}\Big). 
\eqno(4.9)
$$
The heat capacity may be defined by 
$$
C=T\Big({\partial S\over \partial T}\Big)_{\cal N}, \eqno(4.9)
$$
which must be calculated numerically. 

Shown in Fig.1 is a plot of the heat capacity (in units of ${\cal N}k_B$) 
versus temperature (in units of the critical temperature $T_c$) for the 
case of ${\cal N}=2\times 10^3$ and ${\cal N}=2\times 10^6$ particles. 
We choose, for experimental interest[12], the frequency 
eigenvalues $(\omega_1 /2\pi)=(\omega_2/2\pi )=320Hz$, and 
$(\omega_3/2\pi )=18Hz$. 
\begin{figure}[htbp]
\begin{center}
\hspace*{-5mm}\mbox{\epsfig{file=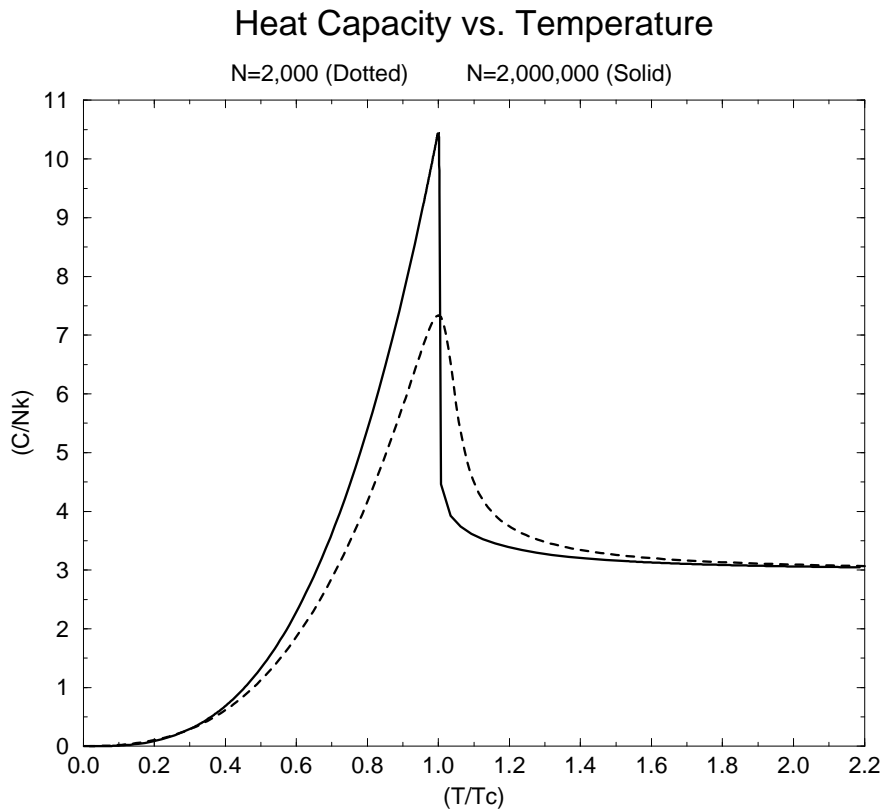, height=14.0cm, angle=-90}}
\end{center}
\caption{The heat capacity (in units of ${\cal N}k_B$) is plotted as a
function of temperature (in units of $T_c$) for ${\cal N}=2\times 10^3$ 
atoms (dotted curve) and ${\cal N}=2\times 10^6$ atoms (solid curve).}
\label{fig: 1}
\end{figure}

\centerline{\bf Figure 1.}
For finite ${\cal N}$, there is strictly speaking 
no Bose-Einstein condensation phase transition. The critical temperature 
$T_c$ is therefore defined as that temperature for which the heat capacity 
reaches the maximum value $C_{max}=C(T_c)$. Although phase transitions 
are defined in mathematics only in the thermodynamic limit 
${\cal N}\to \infty $, for all practical purposes, a quasi-classical 
approximation of 
Eq.(4.1) in the form  
$$
\Xi(T,\mu )=k_BT \int \int \Big({d^3rd^3p\over (2\pi \hbar)^3}\Big)
ln\Big(1-e^{(\mu -h({\bf p},{\bf r}))/k_BT}\Big) 
\ \ (quasi-classical), \eqno(4.10)
$$
where 
$$
h({\bf p},{\bf r})={p^2\over 2M}+
{1\over 2}M{\bf r\cdot }{\hat{\omega }^2}{\bf \cdot r}, \eqno(4.11)
$$       
does yield a Bose-Einstein condensation  phase transition whose heat 
capacity is sufficiently accurate for values of ${\cal N}\sim 10^5$ or higher. 
Thus, we regard recent experiments on Bose atoms confined in a magnetic 
bottle to be probing a physical Bose-Einstein ordered phase. Let us 
consider Eq.(4.10) in more detail.

\section{Quasi-Classical Bose Condensation}

In order to evaluate Eq.(4.10) we employ the quasi-classical form[13,14] 
of Eqs.(4.3) and (4.11); i.e.
$$
q(T)=\int \int \Big({d^3rd^3p\over (2\pi \hbar)^3}\Big)
e^{-h({\bf p},{\bf r})/k_BT}=
\prod_{j=1}^3\Big({k_BT\over \hbar \omega_j }\Big)=
\Big({k_BT\over \hbar \bar{\omega }}\Big)^3, \eqno(5.1)
$$
where $\bar{\omega }=(\omega_1\omega_2\omega_3)^{1/3}$. From 
Eqs.(4.4) and (5.1), it follows that Eq.(4.10) evaluates to  
$$
\Xi(T,\mu )=-\hbar \bar{\omega }
\Big({k_BT\over \hbar \bar{\omega }}\Big)^4
\sum_{n=1}^\infty \Big({1\over n^4}\Big)e^{n\mu / k_BT}. \eqno(5.2)
$$
From Eqs.(4.2) and (5.2), the number of particles obeys 
$$
{\cal N}(T,\mu )=\Big({k_BT\over \hbar \bar{\omega }}\Big)^3
\sum_{n=1}^\infty \Big({1\over n^3}\Big)e^{n\mu / k_BT}, 
\ \ (\mu<0). \eqno(5.3)
$$
With the usual definition of the $\zeta $-function 
$$
\zeta (s)=\sum_{n=1}^\infty \Big({1\over n^s}\Big), 
\ \ {\cal R}e(s)>1, \eqno(5.4)
$$
the Bose-Einstein condensation critical temperature is 
$$
T_c=\Big({\hbar \bar{\omega }\over k_B}\Big)
\Big({N\over \zeta(3)}\Big)^{1/3}. \eqno(5.5)
$$
The non-zero number of Bosons (below the critical temperature) in the 
condensate state is given by 
$$
{\cal N}_0(T)={\cal N}\Big\{1-\Big({T\over T_c}\Big)^3\Big\},
\ \ (T<T_c). \eqno(5.6)
$$
Finally, the entropy below the critical temperature 
$$
{\cal S}(T)=4k_B\zeta (4)\Big({k_BT\over \hbar \bar{\omega }}\Big)^3
=Nk_B\Big({4\zeta(4)\over \zeta (3)}\Big)\Big({T\over T_c}\Big)^3, 
\ \ (T<T_c), \eqno(5.7)
$$
obeys the thermodynamic third law $\lim_{(T\to 0)}{\cal S}(T)=0$.
The heat capacity in the Bose-Einstein Condensed phase is then given by 
$$
\Big({C\over {\cal N}k_B}\Big)=
\Big({12\zeta(4)\over \zeta (3)}\Big)\Big({T\over T_c}\Big)^3
\approx 10.81\Big({T\over T_c}\Big)^3, \ \ (T<T_c). \eqno(5.8)
$$

Employing Eqs.(1.2) and (5.8) we find that the temperature uncertainty 
below the critical temperature obeys 
$$
\Big({\delta T\over T}\Big)\approx \Big({0.3\over \sqrt{\cal N}}\Big)
\Big({T_c \over T}\Big)^{3/2},\ \ (T<T_c). \eqno(5.9)
$$ 
Thus, for ${\cal N}\sim 10^5$ one may safely consider the temperature of the 
ordered phase to be well defined in the range $T_c>T>T_{min}$, where 
$T_{min}\sim 0.05\ T_c$. The open question as to whether the ordered phase is 
a superfluid may now be considered. 

\section{Superfluid Fraction of the Bound Boson System}

The notion of a superfluid fraction in an experimental Bose fluid (such as 
liquid $He^4$) may be viewed in the following manner: Suppose that we pour 
the liquid into a very slowly rotating vessel and close it off from the 
environment. The walls of the vessel are at a bath temperature $T$, and the 
vessel itself rigidly rotates at a very small angular velocity 
${\bf \Omega }$. In the ``two-fluid'' model[15,16], the normal part of the 
fluid rotates with a rigid body angular velocity ${\bf \Omega}$, which 
is the same as the angular velocity of the vessel. On the other 
hand, the superfluid part of the fluid does not rotate. 
The superfluid exhibits virtually zero angular momentum for sufficiently small 
${\bf \Omega }$. The total fluid moment of inertia tensor 
$\hat{I}$ is defined by the fluid angular momentum 
${\bf L}=\hat{I}{\bf \cdot \Omega }$ (as ${\bf \Omega }\to {\bf 0}$). 
We here take the limit ${\bf \Omega }\to {\bf 0}$, to avoid questions 
concerning the effects of vortex singularities on the superfluid. 
The normal fluid, which rotates along with the rotating vessel, contributes 
to the fluid moment of inertia. The superfluid, which does not rotate with 
the vessel, does not contribute to the moment of inertia. Thus, the 
{\it geometric} moment of inertia, 
$$
\hat{I}^{geometric}_{ij}=\int d^3r \bar{\rho }({\bf r})
(r^2 \delta_{ij}-r_ir_j), \eqno(6.1) 
$$
where $\bar{\rho }({\bf r})$ is the mean mass density of the fluid (at 
rest), overestimates the {\it physical} moment of inertia eigenvalues 
when the fluid is actually a superfluid. The normal fluid contributes to 
the moment of inertia and the superfluid does not do so in the limit 
${\bf \Omega }\to {\bf 0}$. Below, we consider in detail the moment of 
inertia of the bound Bose gas.

For the bound Bose system, we consider a mesoscopic rotational state[17] with 
a thermal angular velocity ${\bf \Omega}$. The rotational version of 
Eq.(4.2) reads    
$$
d\Xi_{\bf \Omega }=-{\cal S}dT-{\cal N}d\mu -
{\bf L \cdot}d{\bf \Omega }, \eqno(6.2)
$$
where ${\bf L}$ is the bound Boson angular momentum. Eq.(4.11) gets 
replaced by 
$$
h_{\bf \Omega }({\bf p},{\bf r})={p^2\over 2M}+
{1\over 2}M{\bf r\cdot }{\hat{\omega }^2}{\bf \cdot r}-
{\bf \Omega \cdot}({\bf r\times p}), \eqno(6.3)
$$
so that Eq.(5.1) now reads
$$
q_{\bf \Omega}(T)=\int \int \Big({d^3rd^3p\over (2\pi \hbar)^3}\Big)
e^{-h_{\bf \Omega}({\bf p},{\bf r})/k_BT}=
\Big(q(T)\Big/\sqrt{Det(1-\hat{\omega}^{-2}\hat{\Omega}^2)}\Big), 
\eqno(6.4)
$$  
where the matrix $\hat{\omega }^2$ is written in Eq.(4.8) and 
$$
\hat{\Omega }^2=
\pmatrix{ (\Omega_2^2+\Omega_3^2) & -\Omega_1 \Omega_2 & -\Omega_1 \Omega_3 \cr
-\Omega_1 \Omega_2 & (\Omega_1^2+\Omega_3^2) & -\Omega_2 \Omega_3 \cr
-\Omega_1\Omega_3 & -\Omega_2\Omega_3 & (\Omega_1^2+\Omega_2^2)}. 
\eqno(6.5)
$$
From Eqs.(4.4) and (6.4), it follows that 
$$
\Xi_{\bf \Omega }(T,\mu)=
\Big(\Xi(T,\mu)\Big/\sqrt{Det(1-\hat{\omega}^{-2}\hat{\Omega}^2)}\Big), 
\eqno(6.6)
$$ 
The fluid moment of inertia tensor has the matrix elements 
$$
\hat{I}_{ij}=\lim_{{\bf \Omega }\to {\bf 0}}
\Big({\partial L_i\over \partial \Omega_j}\Big)_{T, \mu}
=-\lim_{{\bf \Omega }\to {\bf 0}}
\Big({\partial^2 \Xi_{\bf \Omega }\over 
\partial \Omega_j \partial \Omega_j}\Big)_{T, \mu}. \eqno(6.7)
$$
Eqs.(4.8), (6.5), (6.6) and (6.7) imply (in the unordered phase)
$$
\hat{I}=-\Xi(T,\mu )
\pmatrix{ \Big({1\over \omega_2^2}+{1\over \omega_3^2}\Big) & 0 & 0\cr
0 & \Big({1\over \omega_1^2}+{1\over \omega_3^2}\Big) & 0 \cr 
0 & 0 & \Big({1\over \omega_1^2}+{1\over \omega_2^2}\Big)}
,\ \ (T>T_c). \eqno(6.8)
$$ 
In the unordered phase, obeying Eq.(5.2), one finds that Eq.(6.8) is precisely 
what would be expected from a normal fluid with geometric moment of inertia 
$$
\hat{I}_{ij}=\int d^3r \bar{\rho }({\bf r})
(r^2 \delta_{ij}-r_ir_j), \ \ (T>T_c), \eqno(6.9) 
$$
where $\bar{\rho }({\bf r})$ is the mean mass density of the atoms.

In the ordered phase $(T<T_c)$, the moment of inertia of the particles 
over and above the condensate is given by Eq.(6.8) with $\mu =0$, i.e. 
$$
\hat{I}^{excitation}=\zeta (4)\hbar \bar{\omega }
\Big({k_BT\over \hbar \bar{\omega }}\Big)^4
\pmatrix{ \Big({1\over \omega_2^2}+{1\over \omega_3^2}\Big) & 0 & 0\cr
0 & \Big({1\over \omega_1^2}+{1\over \omega_3^2}\Big) & 0 \cr 
0 & 0 & \Big({1\over \omega_1^2}+{1\over \omega_2^2}\Big)}
,\ \ (T<T_c). \eqno(6.10)
$$

The question of superfluidity concerns the magnitude of the moment of 
inertia of those particles within the condensate. For $T<T_c$, we use the 
notation that $\hat{I}$ denotes the moment of inertia of the excited Bosons, 
and $\hat{J}$ represents the moment of inertia of the Bose condensate.
If the moment of inertia of the particles in the condensate were zero, 
then the condensate particles would all be ``superfluid''. 

Let $\psi_0 ({\bf r})$ be the normalized ($\int d^3r|\psi_0({\bf r})|^2=1$) 
Bose condensation state. From the geometric viewpoint, the moment of 
inertia of the condensate would be given by 
$$
J^{geometric}_{ij}={\cal N}_0\int d^3r 
|\psi_0 ({\bf r})|^2(r^2 \delta_{ij}-r_ir_j); \eqno(6.11)
$$
i.e. 
$$
\hat{J}^{geometric}={\hbar {\cal N}_0\over 2}
\pmatrix{\Big({1\over \omega_2 }+{1\over \omega_3 }\Big) & 0 & 0 \cr
0 & \Big({1\over \omega_1 }+{1\over \omega_3 }\Big) & 0 \cr 
0 & 0 & \Big({1\over \omega_1 }+{1\over \omega_2 }\Big)}. \eqno(6.12)
$$
The {\it physical} Bose condensate moment of inertia tensor is in reality 
$$
J^{physical}_{ij}={\cal N}_0\sum_\kappa \Big({
<\psi_0|l_i|\psi_\kappa ><\psi_\kappa|l_j|\psi_0>+
<\psi_0|l_j|\psi_\kappa ><\psi_\kappa|l_i|\psi_0>
\over \epsilon_\kappa -\epsilon_0
}\Big), \eqno(6.13)
$$
where ${\bf l}=-i\hbar ({\bf r\times \nabla})$. One may derive Eq.(6.13) by 
treating the rotational coupling $\Delta h=-{\bf \Omega \cdot l}$ to second 
order perturbation theory in the energy $\Delta \epsilon_0({\bf \Omega})$ 
as ${\bf \Omega }\to {\bf 0}$. Eq.(6.13) evaluates to 
$$
\hat{J}^{physical}={\hbar {\cal N}_0\over 2}
\pmatrix{
\Big({(\omega_2-\omega_3)^2\over \omega_2\omega_3(\omega_2+\omega_3)}\Big) 
 & 0 & 0 \cr
 0 & 
 \Big({(\omega_1-\omega_3)^2\over \omega_1\omega_3(\omega_1+\omega_3)}\Big)
 & 0 \cr
 0 & 0 & 
\Big({(\omega_1-\omega_2)^2\over \omega_1\omega_2(\omega_1+\omega_2)}\Big)}. 
\eqno(6.14)
$$

If there exists an axis of symmetry, then the condensate moment of inertia 
corresponding to that axis vanishes. For an experimental example, if 
$\omega_1=\omega_2 \ne \omega_3$, then the $3-$axis is an axis of symmetry 
and $J_{33}^{physical}=0$ as implied by Eq.(6.14). In this sense, for the 
ideal Bose gas, the superfluid fraction is the same as the condensate 
fraction in Eq.(5.6),
$$
\eta_{superfluid} (T)=
\Big\{1-\Big({T\over T_c}\Big)^3\Big\}. \eqno(6.15)
$$  
If we were to employ an axis which is not a symmetry axis, i.e. if the 
angular momentum about that axis were not conserved, then even the 
superfluid would give some contribution to the moment of inertia.

Finally, including the scattering length will have some effect on the 
magnitude of the superfluid fraction[18].

\section{Conclusions}

We have employed the heat capacity $C$ in the relationship 
$$
\Big({\delta T\over T}\Big)\approx \sqrt{\Big({k_B\over C}\Big)} \eqno(7.1)
$$
in order to place lower bounds on possible system temperatures. 
The minimum system temperatures were estimated based on the notion that 
the temperature is ``well defined'' only if temperature fluctuations are 
small $\delta T<<T$. The third law of thermodynamics, in the form 
$\lim_{(T\to 0)}C=0$, dictates that the condition  $\delta T<<T$ is harder 
to achieve as the temperature is lowered. 

For example, one may achieve low temperatures by adiabatic 
demagnetization[19]. The thermodynamics of the method is well 
illustrated by a system of ${\cal N}$ non-interacting two level 
particles, each with possible energies 
$$
E_\pm =\pm \Delta .\eqno(7.2)
$$
Such a physical system might consist of ${\cal N}$ nuclear 
one half spins in a magnetic field. The mean energy of such a system, 
$$
{\cal E}=-{\cal N}\Delta \tanh(\Delta / k_BT), \eqno(7.3)
$$ 
implies a heat capacity $C=(\partial{\cal E}/\partial T)$ given by 
$$
C={\cal N}k_B\Big({(\Delta /k_BT)\over \cosh(\Delta /k_BT)}\Big)^2. 
\eqno(7.4)
$$
The temperature fluctuation of a system of non-interacting two level 
particles follows from Eqs.(7.1) and (7.4) to be 
$$
\Big({\delta T\over T}\Big)\approx \Big({1\over \sqrt{\cal N}}\Big)
\Big({\cosh(\Delta /k_BT)\over (\Delta /k_BT)}\Big). \eqno(7.5)
$$

Thus, for a system of ${\cal N}\sim 10^6$ two level particles, it is only 
possible to achieve a temperature as low as $T_{min}\sim 0.2\ (\Delta/k_B)$. 
For a system of ${\cal N}\sim 10^6$ trapped Bosons, as shown in this work, 
it is possible to achieve a temperature as low as $T_{min}\sim 0.05\ T_c$. 
It is evident on the grounds of temperature fluctuations alone that the 
bound boson systems are the more likely example for the very lowest 
temperatures. Indeed, this has turned out to be true in the laboratory. 

\medskip
\centerline{\bf REFERENCES}
\medskip

\par \noindent 
[1] A. Kent, {\it Experimental Low Temperature Physics}, pp 125-175, 
American Institute for Physics, New York (1993). 
\par \noindent
[2] L. D. Landau and E. M. Lifshitz, {\it Statistical Physics} (Part 1),
p 340, Eq.(112.6), Pergamon Press, Oxford (1980). 
\par \noindent
[3] I. M. Milne-Thompson, {\it The Calculus of Finite Differences}, p 42,
Chelsea Publishing Company, New York (1981).
\par \noindent
[4] C. Jordan, {\it Calculus of Finite Differences}, p 105,
Chelsea Publishing Company, New York (1965).
\par \noindent
[5] L. D. Landau and E. M. Lifshitz, {\it op. cit.}, p 186, Eq.(63.15).
\par \noindent
[6] L. D. Landau and E. M. Lifshitz, {\it op. cit.}, p 160, Eq.(54.4).
\par \noindent
[7] K. Kirsten and D. J. Toms, {\it Phys. Rev.} {\bf A54}, 4188 (1996). 
\par \noindent
[8] H. Haugerud and F. Ravndal, cond-matt/9509041.
\par \noindent
[9] M. H. Anderson, J. R. Ensher, M. R. Matthews, C. E. Wieman and 
E. A. Cornell, {\it Science} {\bf 269}, 198 (1995).
\par \noindent
[10] C. C. Bradley, C. A. Sackett, J. J. Tollett and R. G. Hulet, 
{\it Phys. Rev. Lett.} {\bf 75}, 1687 (1995). 
\par \noindent
[11] K. B. Davis, M.-O. Mewes, M. R. Andrews, N. J. van Druten, 
D. S. Durfee, D. M. Kurn and W. Ketterle, 
{\it Phys. Rev. Lett.} {\bf 75}, 3969 (1995).
\par \noindent
[12] M.-O. Mewes, M. R. Andrews, N. J. van Druten, D. M. Kurn, D. S. Durfee  
and W. Ketterle,  {\it Phys. Rev. Lett.} {\bf 77}, 416 (1996).
\par \noindent
[13] T. T. Chou, Chen Ning Yang, L. H. Yu, {\it Phys. Rev.} {\bf A53}, 4257, 
(1996).
\par \noindent 
[14] T. T. Chou, Chen Ning Yang, L. H. Yu, {\it Phys. Rev.} {\bf A55}, 1179, 
(1997).
\par \noindent 
[15] L. Tisza, {\it Phys. Rev.} {\bf 75}, 885 (1949). 
\par \noindent 
[16] L. D. Landau, {\it Phys. Rev.} {\bf 75}, 884 (1949).
\par \noindent
[17] A. Widom, {\it Phys. Rev.} {\bf 168}, 150 (1968).
\par \noindent
[18] T. T. Chou, Chen Ning Yang, L. H. Yu, {\it op. cit.}
\par \noindent
[19] A. Kent, {\it op. cit.} Chapter 6.

\end{document}